\documentclass{article}
\usepackage{piner}
\usepackage[figuresright]{rotating}
\begin{document}

\submitted{To appear in ApJ 1 April 2005, v622, issue 2}
\title{VLBA Polarization Observations of Markarian 421 After a Gamma-Ray High State}

\author{B. Glenn Piner\altaffilmark{1,2} and Philip G. Edwards\altaffilmark{3}}

\altaffiltext{1}{Department of Physics and Astronomy, Whittier College,
13406 E. Philadelphia Street, Whittier, CA 90608; gpiner@whittier.edu}

\altaffiltext{2}{Visiting Professor, Jet Propulsion Laboratory,
California Institute of Technology, 4800 Oak Grove Drive, Pasadena, CA
91109}

\altaffiltext{3}{Institute of Space and Astronautical Science, Japan Aerospace Exploration Agency
Yoshinodai, Sagamihara, Kanagawa 229-8510, Japan;
pge@vsop.isas.jaxa.jp}

\begin{abstract}
We present four high dynamic range, dual-circular polarization, Very Long Baseline Array (VLBA) observations
at 22 GHz of Markarian 421, taken throughout the year following the source's unprecedented gamma-ray high state
in early 2001.  
Previous VLBI observations of this source had shown only subluminal apparent motions in the
parsec-scale jet, and no apparent connection between jet components and gamma-ray flares, and
so we examined whether the larger gamma-ray flares of 2001 had produced
a component that could be followed on the parsec-scale VLBA images.
These four new VLBA observations are combined with data from our earlier 1999 paper and archival VLBA data-sets that have
become available since 1999 to produce a combined 28 epoch VLBA data-set on Mrk~421 spanning the years 1994 to 2002.
No new component associated with the 2001 flares
was seen on the total intensity images, but the combined data-set allowed precise measurements
of the apparent speeds of the existing components.
The peak measured apparent speed was for component C5, which has an apparent speed of $0.1\pm0.02~c$
($H_{0}=71$ km s$^{-1}$ Mpc$^{-1}$, $\Omega_{m}=0.27$, and $\Omega_{\Lambda}=0.73$).  No counterjet is seen
with a limit on the jet to counterjet brightness ratio $J>\sim100$.
These observed VLBI properties of Markarian 421 are consistent with a jet with a bulk Lorentz factor
$\Gamma\sim2$ and an angle to the line-of-sight $\theta\sim1\arcdeg$, suggesting a jet that decelerates
between the gamma-ray producing region and the parsec scale.
Although a limb-brightened structure is seen in some transverse slices across the jet, it is not seen
consistently, inhibiting interpretation in terms of the fast-spine/slow-layer model that has been invoked
for other sources. 
The VLBI core and inner jet (component C7) have fractional polarizations of $\sim5\%$, and an
electric vector position angle (EVPA) aligned with the jet axis.
Component C5 (at 1.5 mas from the core) has a higher fractional polarization of $\sim15\%$, and 
an EVPA nearly orthogonal to the jet axis.
Significant variability is detected in the EVPA of component C6, which at two of the four epochs shows an
EVPA aligned with the jet axis, possibly a sign of propagating disturbances that are only visible
on the polarization images.  If these propagating disturbances are linked to the 2001 gamma-ray high state,
then their inferred apparent speed is between 1 and $3~c$.
\end{abstract}


\section{Introduction}
\label{intro}
The blazar Markarian 421 has been intensively studied across the electromagnetic spectrum since
its detection as the first known extragalactic source of very high-energy (TeV) gamma-rays (Punch et al. 1992).
Since its initial detection, it has remained among the brightest sources of TeV gamma-rays in the sky,
and has shown multiple episodes of dramatic flaring activity with rapid variability times. 
Notable among these episodes are the dramatic flares of 1996 May, where a flux 10 times that of the
Crab Nebula was recorded (Gaidos et al. 1996; Krennrich et al. 1999), and the
exceptionally strong and long-lasting
flaring activity between 2001 January and March, with a peak flux of 13 times
that of the Crab Nebula (Krennrich, et al. 2001; 2002).

With the confirmed detection of five other blazars in TeV gamma-rays (see, e.g., the review by Horan \& Weekes 2004),
Mrk~421 is now recognized as one of
a class of low-redshift high-energy peaked BL Lac objects that produce significant high-energy gamma-ray 
emission.  The spectral energy distribution of these objects shows two peaks;
in the most common model the low-frequency peak is due to synchrotron
radiation from relativistic electrons in the jet, and the high-frequency peak is due to inverse-Compton scattering.
The flares are produced by shocks in the jet (e.g., Tanihata et al. 2003; Guetta et al. 2004),
with the apparent variability time compressed due to bulk relativistic motion of the jet material.
Homogeneous synchrotron self-Compton (SSC) models of the spectral energy distribution and variability of Mrk~421 require 
very high relativistic Doppler factors in order to reproduce the observed spectral and variability properties.
For example, Maraschi et al. (1999) model the emission from Mrk~421 in 1998 April with $\delta=20$, and 
both Krawczynski et al. (2001) and Konopelko et al. (2003) find
that the emission during 2000 is best modeled by a Doppler factor $\delta\sim50$
(demanding a bulk Lorentz factor $\Gamma>25$).

Imaging the radio emission from relativistic jets with VLBI provides a unique opportunity to study the
evolution of shocks on parsec scales.
In many radio-loud sources, these shocks become visible as superluminally moving 'blobs' or 'components' on VLBI images,
and can sometimes be tracked for years following the flare event with which they are associated (Jorstad et al. 2001b).
The characteristics of these components measured from the VLBI images can provide important constraints
on the properties of the jet.

It came as a surprise when a set of VLBI observations of Mrk~421 spanning 15 epochs from 1994 through 1997
(Piner et al. 1999, hereafter Paper~I) showed only subluminal apparent motions of $0.3~c$
in the jet of Mrk~421 at distances of about 1 pc (projected) from the radio core,
and saw no evidence for a component that could be associated with the 1996 flares.
Low apparent speeds in their parsec-scale radio jets turned out to be a property common to the other TeV blazars
(Edwards \& Piner 2002; Piner \& Edwards 2004), and this property together with other characteristics 
like relatively low radio-core brightness temperatures ($\sim10^{11}$ K),
suggests that the bulk Lorentz factor in the parsec-scale radio
jets is only moderately relativistic.  This can be reconciled with the requirement for high bulk Lorentz
factors on smaller scales by invoking bulk deceleration of the jet along its length (Georganopoulos \& Kazanas 2003),
or by invoking velocity variations perpendicular
to the jet axis such that the radio images are dominated by a lower-Lorentz factor 'layer'
(Giroletti et al. 2004a; Ghisellini, Tavecchio, \& Chiaberge 2004).

Following the exceptionally strong TeV flaring activity in 2001, we decided to observe Markarian 421 again
at four epochs at high-resolution with the NRAO Very Long Baseline Array (VLBA)
\footnote{The National Radio Astronomy Observatory is a facility of the National
Science Foundation operated under cooperative agreement by Associated Universities, Inc.}, to see 
if this record-setting level of high-energy activity would produce a component that could be followed into
the parsec-scale radio jet.  For these observations, we observed with the VLBA in dual-circular polarization mode 
(which provides the intensity and position angle of the polarized flux), so that we could follow
any changes in the magnetic field structure of the jet.
We note that a radio flare at GHz frequencies in which the radio flux density of Mrk~421 increased by a
factor of about 1.5 was also observed in 2001 February, coincident with the increased level of high-energy activity
(Katarzy\'{n}ski, Sol, \& Kus 2003).
Our four epochs of VLBA observations began in 2001 August, about five months after the end of the radio flare.
In addition to the four new epochs of data reported here,
we have also added to our study the archival VLBI data that have become available since Paper~I, to bring the total number
of VLBI observations considered to 28 epochs spanning the years 1994 to 2002.

In this paper we use the cosmological parameters measured by the Wilkinson Microwave Anisotropy Probe (WMAP) of
$H_{0}=71$ km s$^{-1}$ Mpc$^{-1}$, $\Omega_{m}=0.27$, and $\Omega_{\Lambda}=0.73$
(Bennett et al. 2003). 
When results from other papers are quoted,
they have been converted to the set of cosmological parameters given above.
At the distance of Mrk~421 (126 Mpc, z=0.03), 1 milliarcsecond (mas) corresponds to a linear distance of 0.59 pc,
and a proper motion of 1 mas/yr corresponds to an apparent speed of $2.0~c$.
In $\S$~\ref{obs} we discuss the VLBI observations and the results obtained
from imaging and model fitting the VLBI data, and in $\S$~\ref{results} we discuss the astrophysical implications of these results.

\section{Observations}
\label{obs}
\subsection{Details of Observations}
We observed Mrk~421 at four epochs during 2001 and 2002 (2001 Aug 12, 2001 Nov 8, 2002 Jan 26, and 2002 Jul 2)
at 22 GHz with the VLBA,
under observation code BE023.
All of the observations used a standard VLBA continuum setup
(2 intermediate frequencies, each of 8 MHz bandwidth, with 2-bit sampling), and all recorded dual-circular polarization.
The observations averaged 7 hours of time on-source on Mrk~421.
A more detailed observation log is given in Table~\ref{imtab}.

\begin{table*}
\caption{Observation Log and Parameters of the Images}
\label{imtab}
{\tiny \begin{tabular}{l c l c c r c c c c} \tableline \tableline
& Time on & & \multicolumn{3}{c}{Beam Parameters$^{c}$} & Peak Flux & Lowest I & Lowest P & Lowest F \\ 
& Source  & \multicolumn{1}{c}{VLBA} & \multicolumn{1}{c}{Maj.} & \multicolumn{1}{c}{Min.} & \multicolumn{1}{c}{P.A.}
& Density & Contour$^{d}$ 
& Tick$^{e}$ & Pixel$^{f}$ \\ 
Epoch\tablenotemark{a} & (hours) & \multicolumn{1}{c}{Antennas$^{b}$} 
& \multicolumn{1}{c}{(mas)} & \multicolumn{1}{c}{(mas)} & \multicolumn{1}{c}{(deg)} &
(mJy beam$^{-1}$) & (mJy beam$^{-1}$) & (mJy beam$^{-1}$) & (mJy beam$^{-1}$) \\ \tableline \\
2001 Aug 12 & 8 & All   & 0.58 & 0.36 & $-$1.3  & 271 & 0.70 & 0.80 & 1.4 \\ [5pt]
2001 Nov 8  & 8 & No FD & 0.62 & 0.37 & $-$10.6 & 262 & 0.51 & 0.60 & 1.5 \\ [5pt]
2002 Jan 26 & 6 & All   & 0.71 & 0.45 & $-$10.6 & 267 & 0.41 & 0.50 & 1.0 \\ [5pt]
2002 Jul 2  & 5 & No LA & 0.67 & 0.32 & 1.4     & 250 & 0.70 & 0.82 & 2.8 \\ \tableline \\
\end{tabular}}
\\ 
{\scriptsize $a$: The VLBA observation codes associated with these four epochs were
BE023A, BE023B, BE023C, and BE023E, respectively.\\ 
BE023D was not correlated due to technical problems, and was
reobserved as BE023E.}\\ 
{\scriptsize $b$: FD = Fort Davis, Texas; LA = Los Alamos, New Mexico.}\\ 
{\scriptsize $c$: Numbers given for the beam are the FWHMs of the major
and minor axes in mas, and the position angle of the major axis \\ 
in degrees.
Position angle is measured from north through east.
The images have been restored with uvweight=0,$-$1.}\\ 
{\scriptsize $d$: The lowest contour is set to be three times the rms noise
in the I image. Successive contours are each a factor of 2 higher.}\\ 
{\scriptsize $e$: EVPA vectors are drawn at pixels where the polarized flux is
over three times the rms noise in the polarization image.}\\ 
{\scriptsize $f$: Fractional polarization has been displayed only at pixels where the polarized flux is greater than
this value.}
\end{table*}

Calibration, fringe-fitting, detection of cross-polarized fringes, and determination of
the effective polarization response of the feeds (the so-called `D-terms')
were done with standard routines from
the AIPS software package.  
The required electric vector position angle (EVPA) correction
was determined by comparing the observed overall EVPAs of our four calibrator sources (3C~273,
3C~279, OJ~287, and 1156+295) with the EVPAs recorded for these sources on
the VLA/VLBA Polarization Calibration Page (www.aoc.nrao.edu/$\sim$smyers/calibration/),
interpolated to our epochs of observation.
The average EVPA correction derived from the four calibrator sources at each epoch was applied to the Mrk~421 data
using the task CLCOR.
No corrections have been made for Faraday rotation, but because
the cores and jets of BL Lac objects have rotation measures that are
typically 500 radians per meter squared or less (Zavala \& Taylor 2003; 2004),
this will cause a maximum error in the EVPA of 5$\arcdeg$ at 22 GHz.
Self-calibrated CLEAN images from these data-sets were produced using standard CLEAN and
self-calibration procedures from the Difmap software package (Shepherd, Pearson, \& Taylor 1994).
Final polarization images were made using FITSPLOT scripts
(personal.denison.edu/$\sim$homand/FITSplot.v202.tar.gz).
The size of the VLBA beam at 22 GHz is approximately 0.5 milliarcseconds (mas), and this corresponds to a linear resolution
of about 0.3 pc at the distance of Mrk~421.

\subsection{Images}
\label{images}
The 22 GHz VLBA images of Mrk~421 are shown in Figure 1.  
The contours display the total intensity, with the lowest contours
set to three times the rms noise in the I images, and each successive contour a factor of two higher (see Table~\ref{imtab}).
The images have been restored with natural weighting to improve sensitivity to faint jet emission 
(uvweight$=0,-1$ in Difmap). The tick marks show the magnitude of the
polarized flux (with a fixed scale of 25 mas Jy$^{-1}$) and the direction of the EVPA.  
Tick marks are drawn at pixels where the polarized flux is greater than three times the rms noise in the PPOL image.
The color images show the fractional polarization at each pixel, from 0 to 15\%.
Locations of circular Gaussian components placed by the Difmap modelfit task are marked by diamonds, with the identifications
indicated next to the components (see $\S$~\ref{modelfits}).
Numerical parameters for the images in Figure 1 are given in Table~\ref{imtab}.

\begin{figure*}
\plotone{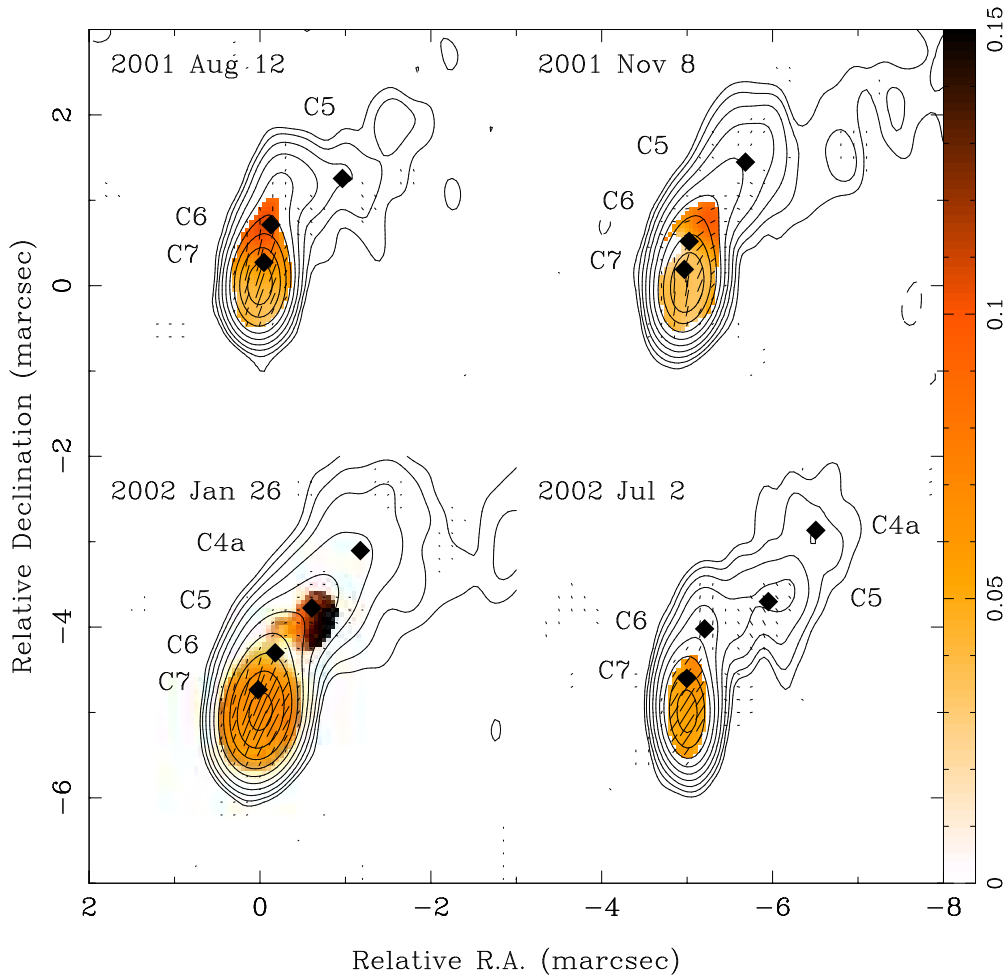}
\caption
{VLBA images of Mrk 421 at 22 GHz.  Contours show the total intensity; the lowest contours are
set to three times the rms noise in the I images, and each successive contour is a factor of two higher.
Tick marks show the magnitude of the
polarized flux (with a fixed scale of 25 mas Jy$^{-1}$) and the direction of the EVPA.
Tick marks are drawn at pixels where the polarized flux is greater than three times the rms noise in the PPOL image.
The color images show the fractional polarization at each pixel.
Locations of circular Gaussian components from model fitting are marked by diamonds, with the component identification
indicated next to the components.  The linear scale is 0.59 pc mas$^{-1}$.}
\end{figure*}

The jet morphology shown in Figure 1 is familiar from earlier papers (Paper~I), so we show only the inner
jet regions on these images.  Component C4 from Paper~I is still detected at about 5 mas from the core, but is
off the scale on the Figure 1 images.  Beyond 5 mas from the core, the surface brightness of the jet 
at 22 GHz falls rapidly
below our detection limit, and no emission is detected beyond about 7 mas from the core.
No counterjet is detected in this source, and from the observations here and in Paper~I we can place
a limit on the jet to counterjet brightness ratio of $J>\sim100$. 

\subsection{Model Fits}
\label{modelfits}
In order to quantify any possible motions in the jet, circular Gaussian model components
were fit to the visibility data at these four epochs using the ``modelfit'' task in Difmap.
The parameters of these four model fits are given in Table~\ref{mfittab}.
In all cases, the reduced chi-squared for the model fit was under 1.0.
The component identification is also given in Table~\ref{mfittab}, using the same naming convention
that was used in Paper~I (some minor exceptions to this are noted in $\S$~\ref{appspeed}).
The identification of components was not based exclusively on these four epochs, but was instead
based on the combined 28 epochs of VLBA data from 1994 to 2002 that is now available on Mrk~421.  
This combined data-set is discussed further in $\S$~\ref{appspeed}.

\begin{table*}[!t]
\begin{center}
{\small
\caption{Circular Gaussian Models}
\label{mfittab}
\begin{tabular}{l c c c r c c c} \tableline \tableline \\ [-5pt]
& & $S^{a}$ & $r^{b}$ &
\multicolumn{1}{c}{PA$^{b}$} & $a^{c}$ & & $T_{B}^{e}$ \\ [5pt]
Epoch & Component & (mJy) & (mas) &
\multicolumn{1}{c}{(deg)} & (mas) & $\chi_{R}^{2}$\tablenotemark{d} & ($10^{11}$ K) \\ \tableline \\ [-5pt]
2001 Aug 12 & Core & 249  & ...  & ...     & 0.06 & 0.94 & 1.6 \\ [5pt]
            & C7   & 58   & 0.29 & $-$10.0 & 0.19 &      &     \\ [5pt]
            & C6   & 19   & 0.74 & $-$7.5  & 0.24 &      &     \\ [5pt]
            & ...  & 17   & 1.23 & $-$12.5 & 0.33 &      &     \\ [5pt]
            & C5   & 16   & 1.60 & $-$37.3 & 0.99 &      &     \\ [5pt]
            & C4   & 12   & 4.52 & $-$43.3 & 2.77 &      &     \\ [5pt]
2001 Nov 8  & Core & 225  & ...  & ...     & 0.04 & 0.92 & 3.4 \\ [5pt]
            & C7   & 52   & 0.22 & 9.4     & 0.17 &      &     \\ [5pt]
            & C6   & 44   & 0.55 & $-$2.1  & 0.25 &      &     \\ [5pt]
            & ...  & 17   & 0.95 & $-$14.8 & 0.28 &      &     \\ [5pt]
            & C5   & 22   & 1.62 & $-$24.7 & 0.65 &      &     \\ [5pt]
            & C4   & 18   & 4.79 & $-$45.2 & 2.20 &      &     \\ [5pt]
2002 Jan 26 & Core & 245  & ...  & ...     & 0.08 & 0.72 & 0.9 \\ [5pt]
            & C7   & 43   & 0.29 & 5.8     & 0.17 &      &     \\ [5pt]
            & C6   & 34   & 0.74 & $-$9.2  & 0.32 &      &     \\ [5pt]
            & C5   & 14   & 1.39 & $-$25.5 & 0.43 &      &     \\ [5pt]
            & C4a  & 15   & 2.25 & $-$31.2 & 1.01 &      &     \\ [5pt]
            & C4   & 13   & 5.07 & $-$44.0 & 1.31 &      &     \\ [5pt]
2002 Jul 2  & Core & 245  & ...  & ...     & 0.08 & 0.73 & 0.9 \\ [5pt]
            & C7   & 48   & 0.42 & 0.3     & 0.24 &      &     \\ [5pt]
            & C6   & 24   & 1.02 & $-$11.7 & 0.37 &      &     \\ [5pt]
            & C5   & 20   & 1.62 & $-$35.9 & 0.65 &      &     \\ [5pt]
            & C4a  & 8    & 2.63 & $-$35.1 & 0.86 &      &     \\ [5pt]
            & C4   & 15   & 5.77 & $-$44.3 & 1.90 &      &     \\ \tableline \\
\end{tabular}}
\end{center}
\vspace{-0.15in}
$a$: Flux density in millijanskys. \\
$b$: $r$ and PA are the polar coordinates of the
center of the component relative to the presumed core.
Position angle is measured from north through east. \\
$c$: $a$ is the Full Width at Half Maximum (FWHM) of the circular Gaussian
component. \\
$d$: The reduced chi-squared of the model fit. \\
$e$: Source frame, homogeneous sphere brightness temperature of the core component
(Hirabayashi et al. 1998 and erratum).
\end{table*}

Note in Figure~1 that the model components are not well-defined on the images as individual blobs
separated by gaps on the contour maps.  Instead, the model components appear as local maxima in the jet
emission, or as `shoulders' where the jet emission stays roughly constant for a short distance.
However, the model-fitting algorithm we have used fits components to the observed visibilities, not to
the image, and sub-beam resolution is obtained if the signal-to-noise ratio is high.
The remarkable consistency of the model-fit positions obtained from epoch to epoch, over all 28 epochs of data,
provides confidence in the robustness of the identifications in Table~\ref{mfittab}.

\section {Results and Discussion}
\label{results}

\subsection{Apparent Speed}
\label{appspeed}
In Paper~I we combined model fits of all of the archival VLBI data available on Markarian 421 at that time (15 epochs of
data spanning the years 1994 to 1997) to measure the apparent speeds of components in the parsec-scale radio jet.
The fastest speed measured in the jet in Paper~I was $\approx 0.3\pm0.1~c$ for component C5.
In this paper, we improve the measurement of apparent speeds by adding the four new VLBA epochs observed for
this paper, plus all of the archival VLBA observations observed since Paper~I.
The total combined data-set now comprises 28 epochs spanning the years 1994 to 2002; all 
of these epochs were observed with the VLBA.

We have added the following VLBI data to the 15 epochs published in Paper~I:
\begin{enumerate}
\item{The four VLBA observations discussed in this paper, whose model fits are given in Table~\ref{mfittab}.}
\item{Eight new VLBA observations at 8 GHz from the U.S. Naval Observatory's Radio Reference Frame Image Database (RRFID)
\footnote{http://rorf.usno.navy.mil/RRFID/}, spanning the dates 1997 March through 2002 January.
The self-calibrated visibilities for these eight observations were independently modeled by us in Difmap, as part of our
study of RRFID source kinematics (Piner et al. 2003).  Model fitting of the 
8 GHz RRFID observations used in Paper~I was redone, 
so that all of the RRFID observations were modeled in a uniform manner; however,
there were no significant differences between the original fits and the new fits.}
\item{The seven epochs of VLBA data at 15 GHz observed by the 2~cm survey (Kellermann et al. 2004),
spanning the dates 1995 April through 2001 March.
Here we have used the model-fit component positions used by the 2~cm survey
(Ken Kellermann, private communication), but 
based on our additional 21 epochs of data, we have altered the component identifications from those used
by Kellermann et al. (2004).}
\end{enumerate}

Because of the addition of so much high-quality VLBA data on this source since Paper~I, we have removed the
following observations used in Paper~I from our data set:
\begin{enumerate}
\item{All observations from Paper~I not observed by the VLBA, so
that all epochs are observed by the same telescope.  Observations removed include five epochs of Mark III geodetic
VLBI observations, and a single VSOP observation.}
\item{All observations from Paper~I at frequencies at or below 5 GHz, because they do not have the resolution to
properly deconvolve the model components seen at higher frequencies.}
\end{enumerate}

The final data set used for the apparent speed measurement then comprises 28 epochs observed with the VLBA at
frequencies ranging from 8 to 43 GHz.  The 28 epochs consist of 7 epochs from the 2~cm survey, 14 epochs from the
RRFID, and 7 epochs of higher dynamic range full-track observations, including the four new observations presented in this paper. 
The analysis in Paper~I showed negligible frequency-dependence of the separation of components from the core,
justifying the combination of observations at different frequencies. 
This is confirmed with the new data in this paper --- the average difference in the core separation of components
between the 2~cm survey observations and the 8 GHz RRFID observations within one month of each other is 0.05 mas,
in a direction opposite to that expected from the varying optical depth model for frequency-dependent separation
(components farther from the core at higher frequencies).  Since this difference is about a factor of four
smaller than our typical uncertainty for component separations, we conclude that there is no measurable
frequency-dependent separation within the precision of these measurements.

Figure 2 shows the separation of the Gaussian model components from the VLBI core vs. time for this final data set.
The component identifications are listed on the right of this figure, and are the same as those used in Paper~I
with the following exceptions: a faint component occasionally seen between C4 and C5 in Paper~I has now been seen 
enough to define its motion, and it is now included as component C4a, and a region interior to C6 noted as confused
in Paper~I is now modeled as component C7.
Errors on the component separations are assumed to be proportional to the beam size and also to the surface 
brightness of the component, with the centroid of compact inner components determined more accurately
than those of diffuse outer components like C4.
Accordingly, error bars of one-half the beam size 
were assigned to the separations of C4 and C4a from the core, one quarter of the beam
size to the separations of C5 and C6 from the core, and one-eighth of the beam size to 
the separation of C7 from the core
(for beam size we use the extent of the projection of the
two-dimensional beam onto a line joining the center of the core to the center of the Gaussian component).

\begin{figure*}[!t]
\plotfiddle{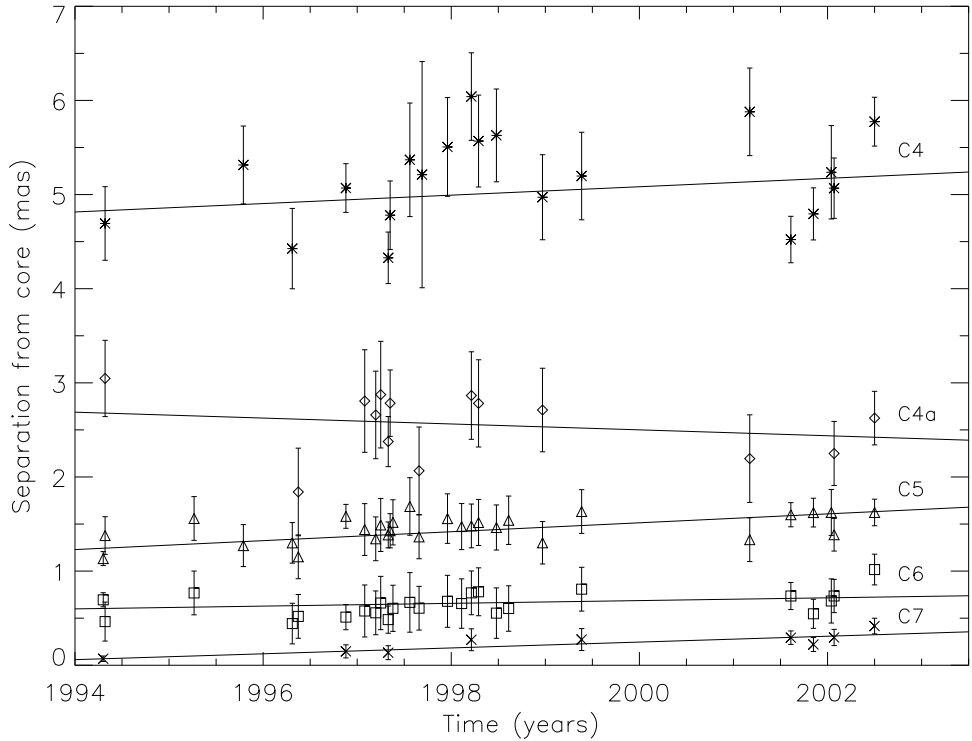}{4.0in}{90}{60}{60}{220}{-30}
\caption{
Distances from the core of Gaussian component centers as
a function of time.  The lines are the linear least-squares fits to outward motion with constant speed.
Asterisks represent component C4, diamonds C4a, triangles C5, squares C6, and crosses C7.}
\end{figure*}

Linear least-squares fits to the component separations vs. time were used to measure the proper motions and apparent
speeds of the components; these fits are indicated by solid lines on Figure~2.
Measured proper motions are $45\pm33$, $-31\pm44$, $47\pm12$, $15\pm13$, and $31\pm6$
$\mu$as yr$^{-1}$ for C4, C4a, C5, C6, and C7 respectively.
Measured apparent speeds are $0.089\pm0.066~c$, $-0.063\pm0.088~c$, $0.095\pm0.024~c$, $0.029\pm0.026~c$, and
$0.062\pm0.013~c$ for C4, C4a, C5, C6, and C7 respectively.
There is no sign of acceleration or deceleration --- second order fits 
were tried but did not have appreciably higher significance than the
linear fits.  
Preliminary measurements of these apparent speeds were given in Table~3 of Piner \& Edwards (2004), the speeds
listed here should now be used in place of the preliminary values quoted in that paper.
Note that the analysis of the seven epochs of data from the 2~cm survey by Kellermann et al. (2004)
used an alternative component identification scheme which resulted in faster, but still subluminal, speed measurements 
(average component speed of $0.4~c$).
With approximately four times the amount of data available, we suggest the alternative identification used in Figure~2,
which the 2~cm survey data points also match well. 
The exceptionally dense time sampling,
particularly during the years 1997 and 1998, would seem to preclude the faster speeds given by Kellermann et al. (2004).

At this point it is also worthwhile revisiting the speculative links between our components
and the Zhang \& B\aa\aa th (1990) (hereafter ZB90) components made in Paper~I.
ZB90 obtained four epochs of VLBI data on Mrk~421 at 5 GHz during the early 1980s and fit this data
with superluminal components.  As discussed in Paper~I, their data can also be fit by    
subluminal components, and their interpretation in terms of superluminal components was forced
by their underestimation of the errors on model component positions (they used 1\% of the beam).
In Paper~I we identified ZB90's Component 2 from their Table 1 with our C4,
and ZB90's Component 3 from their Table 1 with our C3.
An alternative identification was made by Edwards et al. (2000) that matched
ZB90's Component 2 with our C4a, and ZB90's Component 3 with our C4.
Here we confirm that this alternative identification is more plausible, because inclusion
of the additional data on Mrk~421 from this paper effectively rules out the old interpretation from Paper~I.
With the inclusion of the ZB90 data, the fitted proper motions of C4 and C4a are $50\pm14$ and $39\pm15$
$\mu$as yr$^{-1}$ (corresponding to apparent speeds of $0.100\pm0.027~c$ and $0.078\pm0.030~c$)
for C4 and C4a, respectively.
However, since any identification of components across a ten-year gap is always speculative,
in the rest of this paper we use the motions derived in the preceding paragraph without the ZB90 data.

\subsection{Kinematics}
\label{kinematics}
Based on the large increase in the amount of data available for this source, we have revised
the measured apparent speed for the fastest component in Mrk~421 (component C5) since Paper~I.
The measured apparent speed of this component from Paper I was $0.3\pm0.1~c$, the new
apparent speed from this paper is $0.1\pm0.02~c$, about $2\sigma$ below the previous value.
The new value is more reliable due to the use of higher sensitivity, higher resolution data,
and a longer time baseline of monitoring.
In this section, we assume that this apparent speed represents the bulk apparent speed of the jet plasma,
and discuss the implications of this for the true speed and orientation of the jet.
Three other jet components  (C4, C4a, and C6) have measured apparent speeds from our data that are consistent with no motion
(although note that C4 and C4a have measured speeds more consistent with C5 and C7 if the ZB90 data is included).
These features may be stationary components, which are a common feature in radio jets.
Jorstad et al. (2001a) discuss several possible causes of stationary components, including standing recollimation
shocks caused by pressure imbalances between the jet and the external medium, and stationary shocks at the
location of a bend in the jet.  The jet in Mrk~421 does bend continuously from north to west over the range studied in
this paper, but the definitive test for determining the nature of a stationary component is to observe
the behavior of a moving component that passes through the stationary component (Jorstad et al. 2001a),
an event that has not yet occurred in our observations of Mrk~421.

It was noted in Paper~I and by Piner \& Edwards (2004) that the slow apparent speeds in TeV blazars are
difficult to reconcile with the high Doppler factors required to fit the multiwavelength spectra
and variability.  
The problem arises when the equations for apparent speed ($\beta_{app}=\beta\sin\theta/(1-\beta\cos\theta)$,
where $\beta=v/c$ and $\theta$ is the angle of the jet to the line-of-sight)
and Doppler factor ($\delta =1/\Gamma(1-\beta\cos\theta)$, where $\Gamma$ is the bulk Lorentz factor) are
solved simultaneously for the Lorentz factor and viewing angle, then very small viewing angles are obtained.
The current observations of Mrk~421 offer an example of this, using an apparent speed of $0.1~c$
and a Doppler factor of 50 as found in some homogeneous SSC models (Krawczynski et al. 2001; Konopelko et al. 2003),
yields a Lorentz factor of 25 and a viewing angle of 0.005$\arcdeg$.
While less extreme than this example for Mrk~421,
the other TeV blazars also yield viewing angles of $\sim1\arcdeg$ or less when this method is applied
(Piner \& Edwards 2004).  These small viewing angles and high Lorentz factors seem to be in conflict
with those required by unification models of BL Lacs and FR~I radio galaxies (Tavecchio 2004).
This problem has been dubbed the ``$\delta$ crisis'' by Tavecchio (2004).
It can be argued that the TeV emitting BL Lacs are extreme BL Lacs, and that the unification
argument would not apply to a few extreme sources.  However, TeV sources are all, of necessity, at
redshifts $z< \sim0.15$, and there must be many more such objects at higher redshifts.
Lacking a larger number of source counts such as might be provided by the VERITAS telescope, this remains
an open question, but it seems unlikely that the known TeV emitting BL Lacs would all have angles to the
line-of-sight of less than $1\arcdeg$.

One proposed solution to the ``$\delta$ crisis'' is that the jet decelerates along its length,
so that the bulk Lorentz factor in the VLBI portion of the jet at parsec scales is substantially
less than that in the gamma-ray emitting portion at sub-parsec scales (Georganopoulos \& Kazanas 2003). 
This helps in two ways: a smaller Lorentz factor at VLBI scales means the viewing angle can be larger and still
reproduce the observed apparent speed, and since the faster regions of the jet closer to the core see 
beamed seed radiation from the slower regions of the jet farther out,
the same inner-jet emission can be reproduced with a lower Lorentz factor than that required in
homogeneous models.  Georganopoulos \& Kazanas (2003) suggest that the required Lorentz factor and Doppler factor in
the inner jet are about a factor of two lower in this case than in the homogeneous models.
VLBI observations can provide constraints on the viewing angle and the amount of deceleration through 
the measured apparent speed and the jet to counterjet brightness ratio limit. 
For Mrk~421, if we keep the apparent speed fixed at the observed value of $0.1~c$, then a jet with
a viewing angle of $2\arcdeg$ has a Lorentz factor of 1.5 and a jet to counterjet ratio of 100
(using $J=((1+\beta\cos\theta)/(1-\beta\cos\theta))^{2-\alpha}$, where $\alpha=-0.5$ is the spectral index,
see Paper I),
which is approximately equal to the observed lower limit ($\S~\ref{images}$). 
A jet with a viewing angle of $1\arcdeg$ has a Lorentz factor of 2 and a jet to counterjet ratio of 700,
comfortably above the observed lower limit.
Thus a jet that begins with a Lorentz factor $\Gamma\sim15$ in the inner jet region (Georganopoulos \& Kazanas 2003),
and decelerates to $\Gamma\sim2$ on VLBI scales with a viewing angle $\theta\sim1\arcdeg$ would be consistent
with the observed VLBI properties of this source. 

In this decelerating jet model a substantial amount of the jet's bulk kinetic energy has been lost by VLBI scales.
A favored mechanism for the transfer of bulk kinetic energy into particle kinetic energy and eventually to
radiation involves internal shocks in the jet (Spada et al. 2001).
In powerful blazars, such as 3C~279, such models should work with low efficiency, because the jet is known to remain
highly relativistic out to many parsecs from the VLBI core (Wehrle et al. 2001).
We suggest that when such shock models are applied to the TeV blazars in particular
(e.g. Guetta et al. 2004), then they need to work with high efficiency
if the observed VLBI properties of these sources are to be reproduced.
We also note that low parsec-scale Lorentz factors seem to be a property of the HBL class in general,
not just of the TeV blazars, because Giroletti et al. (2004b) find a typical bulk Lorentz factor around $\Gamma=3$
in their study of a sample of about 20 HBLs.

A proposed alternative to a jet that decelerates along its length is a jet with a velocity
structure {\em transverse} to the jet axis: a fast inner `spine' and a slower outer `layer'.
The outer layer has been decelerated by interaction with the external medium, or simply was not accelerated
as efficiently as the spine.
The VLBI observations see predominantly emission from the layer with its low Lorentz factor, while
the high-energy observations see predominantly emission from the high-Lorentz factor spine.
Two variants of this model have recently been proposed for TeV blazars by Giroletti et al. (2004a)
and Ghisellini et al. (2004).
Giroletti et al. (2004a) applied a spine/layer model to their observations of Mrk~501.
In that paper, the difference in the brightness of the spine and the layer was assumed to be due to their changing Doppler
factors as the jet bent away from the line-of-sight, from 5 degrees to 25 degrees.  
Such a model cannot match the VLBI observations of Mrk~421 from this paper, because the slow apparent speed
of $0.1~c$ must be produced by a small angle to the line-of-sight (solutions with $\beta_{app}<\beta$ occur only
for small angles, or for $\theta>90\arcdeg$), where the fast spine would dominate the emission.

A different version of the spine/layer model is presented by Ghisellini et al. (2004).
They assume the angle of the jet to the
line-of-sight is constant, so that the slow layer and fast spine have constant Doppler factors.
Unlike Giroletti et al. (2004a), 
they model the emission from both the layer and spine, and take into account the inverse-Compton radiation  
each portion produces from the seed radiation coming from the other portion.
They find that the slow layer can dominate the emission at radio frequencies over the fast spine,
even when the layer has a smaller Doppler factor --- see, for example, their Figure~3 for Mrk~501.
However, the specific model they present for Mrk~421 has the fast spine dominating at all frequencies.

The simple observational test for the spine/layer model is that, if the VLBI emission is coming predominantly from
the layer, then the jet should appear limb-brightened on the VLBI images, provided the jet
can be resolved in the transverse direction.
Such a limb-brightened structure was seen by Giroletti et al. (2004a) in their VLBI images of Mrk~501.
Figure~3 shows the surface brightness of the Mrk~421 jet vs. the transverse distance from the jet axis,
at distances from the core ranging from 1 to 6 mas, taken from the VLBA image
of 2002 Jan 26 (the image with the highest dynamic range, the other epochs produce similar results). 
Examining these six slices, the jet is still unresolved transversely at 1 mas from core,
at 2 mas the jet begins to be resolved in the transverse direction, and at 3 mas the jet shows an
unambiguous limb-brightened structure.  However, this clear limb-brightened structure is gone in
the following three slices at 4, 5, and 6 mas from the core (there are suggestions of
small dips in the jet's transverse profile, but not at a level where they could be called significant).
We conclude that we do not see consistent evidence for a clear spine/layer structure in the Mrk~421 jet at
distances of several mas from the core.
Possibly the varying character of the transverse structure in Figure~3 is due to instabilities developing 
in the jet, as immediately beyond the region included on this figure the jet loses collimation and becomes
diffuse, as seen in the lower resolution images in Paper~I. 

\begin{figure*}[!t]
\plotone{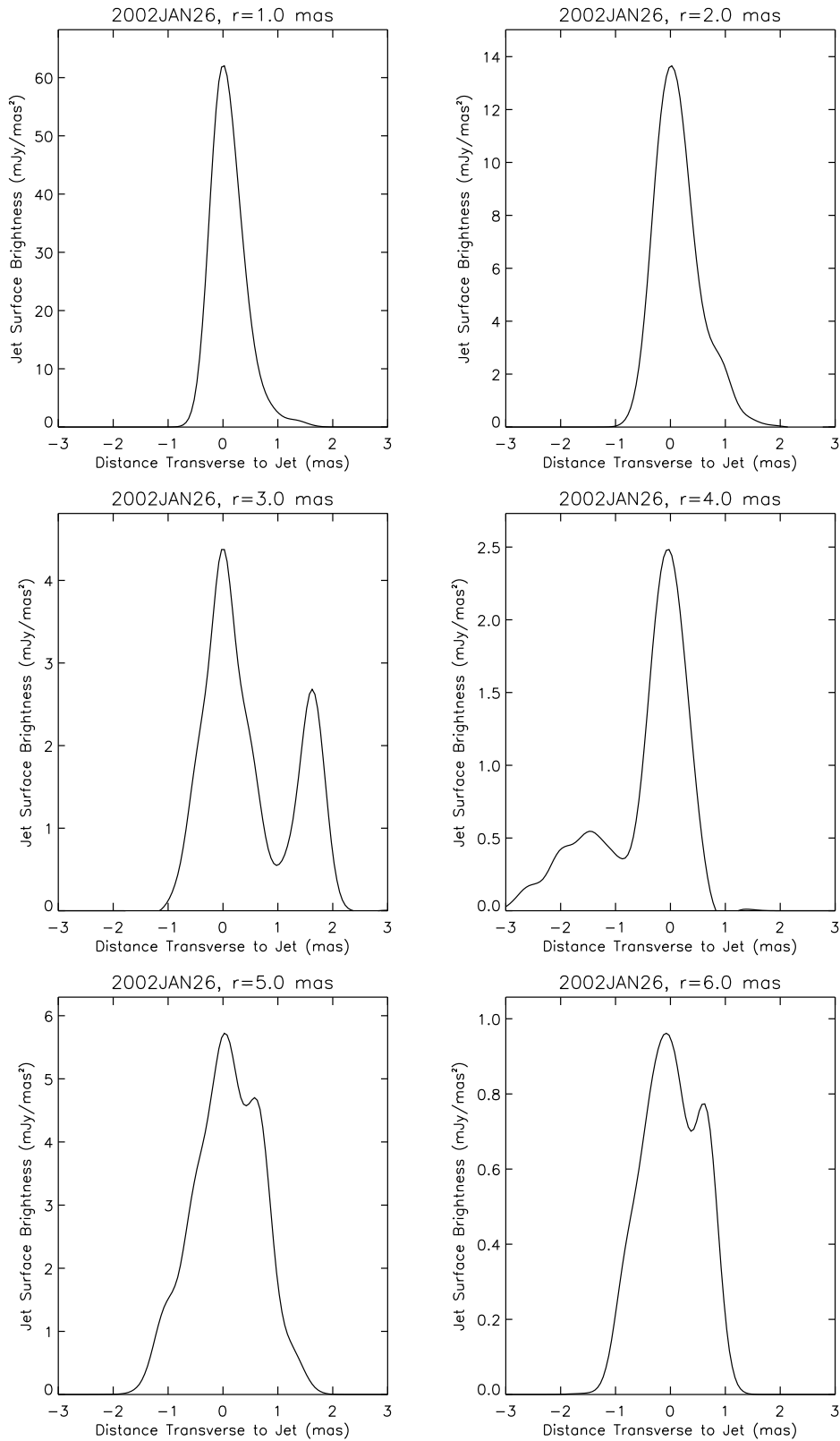}
\caption
{Jet surface brightness versus transverse distance from the jet axis at six
different distances from the core on 2002 Jan 26.}
\end{figure*}

If the measured apparent speeds do not represent the bulk motion of the plasma, but only of
stationary or slowly moving patterns, then the model constraints derived from the apparent
speed would not apply.  In that case, the relevant VLBI observables for constraining the kinematics
would be limited to the brightness temperature and the jet-to-counterjet brightness ratio limit.
The relatively low brightness temperature of Mrk~421 would still constrain the Doppler factor to be small in the
VLBI core, if Mrk~421 has a similar intrinsic brightness temperature to other compact radio sources (Readhead 1994).
In the following section, we investigate whether the propagation of
a disturbance down the Mrk~421 jet might have been seen in the jet's polarization properties,
but not in the total intensity images.

\subsection{Polarization Properties}
VLBI polarization observations of Mrk~421 have recently been presented by
Marscher et al. (2002) and Pollack, Taylor, \& Zavala (2003). 
Both of these papers only detect polarized emission from the VLBI core,
presumably due to the snapshot nature of their observations.
Marscher et al. (2002) find the core of Mrk~421 to be 2\% polarized with an EVPA nearly aligned
with the inner jet position angle, in a 22 GHz image from 1997.
Pollack et al. (2003) also find the VLBI core to be 2\% polarized based on a 
5 GHz image from 2000.  The EVPAs reported by Pollack et al. (2003) are likely to be
affected by large errors due to the lack of Faraday rotation corrections at their lower observation frequency.

Table~\ref{ptab} summarizes the polarization properties of the components in
Mrk~421 at each epoch from our observations.  Only components between C5 and the core are included,
because the outer components are too weak for the detection of polarized emission in these observations.
Table~\ref{ptab} gives the fractional polarization of the component, the electric 
vector position angle (EVPA) of the polarized emission (position angle is measured from north
through east), and the offset of the EVPA from the jet position angle at the location of the component.
Because the least-squares fits of $\S~\ref{appspeed}$ show that the components are effectively
stationary during the 11 month span of these observations (maximum motion is 42 $\mu$as for C5),
we have evaluated the polarization properties using 
the fractional polarization and EVPA of the images at the average position of
the component during the four epochs, in order to minimize errors arising from variations in the model fits.

\begin{table*}[!t]
\begin{center}
{\small
\caption{Polarization Properties}
\label{ptab}
\begin{tabular}{l c r r r} \tableline \tableline \\[-5pt]
& & \multicolumn{1}{c}{$m^{a}$} &
\multicolumn{1}{c}{EVPA$^{b}$} & \multicolumn{1}{c}{EVPA$-$JPA$^{c}$} \\ [5pt]
Epoch & Component & \multicolumn{1}{c}{(\%)} &
\multicolumn{1}{c}{(deg)} & \multicolumn{1}{c}{(deg)} \\ \tableline \\[-5pt]
2001 Aug 12 & Core & $4.4\pm1.3$  & $-$18.7 & $-$20 \\ [5pt]
            & C7   & $6.8\pm1.9$  & $-$16.0 & $-$17 \\ [5pt]
            & C6   & $9.6\pm2.8$  & $-$6.2  & 1     \\ [5pt]
            & C5   & $14.1\pm6.2$ & 42.9    & 74    \\ [5pt]
2001 Nov 8  & Core & $3.7\pm1.0$  & $-$9.2  & $-$11 \\ [5pt]
            & C7   & $3.2\pm0.9$  & $-$17.8 & $-$19 \\ [5pt]
            & C6   & $8.7\pm2.5$  & $-$59.1 & $-$52 \\ [5pt]
            & C5   & $13.7\pm4.6$ & 56.6    & 88    \\ [5pt]
2002 Jan 26 & Core & $6.5\pm1.8$  & $-$23.4 & $-$25 \\ [5pt]
            & C7   & $6.7\pm1.9$  & $-$20.1 & $-$22 \\ [5pt]
            & C6   & $2.9\pm0.8$  & $-$38.8 & $-$31 \\ [5pt]
            & C5   & $13.3\pm4.0$ & 74.0    & 105   \\ [5pt]
2002 Jul 2  & Core & $5.4\pm1.5$  & $-$32.1 & $-$34 \\ [5pt]
            & C7   & $5.9\pm1.7$  & $-$32.5 & $-$34 \\ [5pt]
            & C6   & $7.4\pm2.2$  & $-$23.5 & $-$16 \\ [5pt]
            & C5   & $34.8\pm9.9$ & 33.3    & 64    \\ \tableline
\end{tabular}}
\\ [5pt]
\end{center}
$a$: Fractional polarization of component. \\
$b$: Electric Vector Position Angle of the polarized emission from the component.
A $\pm10\arcdeg$ uncertainty is assumed for the measurement of EVPA (see text). \\
$c$: Offset of the EVPA from the jet position angle
at the location of the component. \\
\end{table*}

To estimate the errors on the quantities in Table~\ref{ptab}, we used the empirical study of polarimetry
errors done by Homan et al. (2002), who compared simultaneous polarization images at 15 and 22 GHz.
Homan et al. (2002) estimated that in the worst cases the measured fluxes of jet components were accurate to
20\%, and the measured EVPAs to 10$\arcdeg$.
We have accordingly used a 20\% flux error (added in quadrature to the rms noise in the images)
to derive the error bars on fractional polarization, and assume 10$\arcdeg$ error bars for the angular
quantities in the last two columns of Table~\ref{ptab}.
We have purposely chosen the most conservative error bars from Homan et al. (2002), in order to make sure
that no spurious variability is detected in the polarization properties.

The emission from the core of Mrk~421 is polarized
with a relatively high weighted-average fractional polarization of 4.6\%.
Similar fractional polarizations are seen for the inner jet components C7 and C6.
The fractional polarization increases in component C5, whose weighted-average fractional polarization
is 15.1\%.  This is in agreement with observations of Lister (2001), who observed that
fractional polarization of VLBI jets initially increases with distance from the core.
The fractional polarization of Mrk~421's core seems to have increased from the lower value of
2\% measured in 1997 and 2000 by Marscher et al. (2002) and Pollack et al. (2003), respectively.

The EVPAs of the core and C7 are similar, and are aligned with the jet position angle
with a typical offset of $\approx20\arcdeg$.
As the emitting material of the core and jet of Mrk~421 is optically thin at 22 GHz (the spectral index
of the VLBI core between 15 and 43 GHz was measured to be $-0.54$ in Paper I,
see also the quasi-simultaneous data of Kovalev et al. [1999]),
the jet rest-frame magnetic field is oriented perpendicular to the EVPA (Aller 1970; see also
Lyutikov, Pariev, \& Gabuzda [2004] for a discussion of inferred directions of rest-frame vs. observer-frame 
magnetic fields).
A magnetic field oriented nearly orthogonal to the jet is characteristic of emission from
transverse shocks. 
This result is also consistent with those of Lister (2001), who found
that cores with high ($>4\%$) VLBI fractional polarization tended to have EVPAs aligned within $30\arcdeg$ 
of the jet position angle.
The EVPA of component C5 is different from that of the core and C7, 
and is nearly orthogonal to the jet (average EVPA$-$JPA$\approx80\arcdeg$),
implying a magnetic field oriented nearly parallel to the jet.
This is unusual for a jet component in a BL Lac object.  Lister (2001), 
Zavala \& Taylor (2003), and Marscher et al. (2002) find no jet components in BL Lac
objects with EVPAs misaligned by more than $\approx45\arcdeg$ from the jet position angle; however,
the samples are quite small in all of these cases.

When we test the quantities in Table~\ref{ptab} for time variability, the only one that
shows variability at greater than 3$\sigma$ significance is the EVPA at the location of
component C6 (0.003 probability of no variation).
The EVPA for this component in 2001 August is aligned parallel to the jet ($B$ perpendicular). 
By 2001 November it has rotated by
$53\arcdeg$ to make an oblique angle with the jet, it then rotates back toward its 2001 August value during
2002 January, becoming aligned parallel to the jet again by 2002 July.
We are confident that this represents a real EVPA variation --- the detections are well above the
noise level, and none of the other components show similar systematic swings.

If this EVPA swing were due to Faraday rotation, it would require a rotation measure variation of 5000 radians m$^{-2}$
between 2001 August and 2001 November.
We consider this scenario unlikely, because the required rotation measure change is an order of magnitude larger
than typical rotation measures in BL Lac objects (Zavala \& Taylor 2003).
One speculation for the behavior of the EVPA at this location is that 
propagating shocks are causing the magnetic field to become aligned perpendicular to the jet 
at this location in 2001 August
and 2002 July, and thus changes in EVPA may provide a way to track shocks that cannot be separated from
the quasi-stationary components on the total intensity images.
If either of these perpendicular alignments of the $B$ field of C6 (in 2001 August and 2002 July) is associated
with a disturbance that left the core at the time of the 2001 gamma-ray high state,
then that disturbance is propagating superluminally with an apparent speed between 1 and $3~c$.
Similar constraints are placed on this propagation speed by assuming the hypothetical shock had traveled
beyond C7 by the first epoch, and not reached C5 by the final epoch.
However, why these shocks would not be visible as moving components on the total intensity images is not known.

\section{Conclusions}
This paper has presented four high dynamic range, dual-circular polarization, 22 GHz VLBA images of Mrk~421
taken during the year following this source's unprecedented gamma-ray high state in early 2001.
The goal of the VLBA observations was to identify and measure the kinematics of any new component
ejected into the parsec-scale jet as a result of this flaring activity.
No new component associated with the 2001 flares
was seen on the total intensity images,
and the structure on these images seems decoupled from the high-energy activity in the core.
Based on this hypothetical components non-appearance by the fourth epoch, we can place an
upper limit to its apparent speed of $<0.3~c$.
However, the non-detection of a new superluminal component still provides valuable constraints
on the energetics of this source, and implies that the jet loses much of its kinetic energy
by the time it reaches parsec scales.

We combined the VLBA data from this paper and Paper I, and added archival VLBA data 
that had become available since Paper I, to
create a 28 epoch set of VLBA data on this source,
in order to measure the kinematics of the parsec-scale jet as precisely as possible.
We improved the apparent speed measurement over that in Paper I by a factor of a few, with the
current highest measurement of the apparent speed being $0.1\pm0.02~c$ (for component C5).
Taken together with the limit on the jet to counterjet brightness ratio ($J>\sim100$), this
suggests a low Lorentz factor jet close to the line-of-sight --- a jet with
a bulk Lorentz factor
$\Gamma\sim2$ and an angle to the line-of-sight $\theta\sim1\arcdeg$ would be consistent
with the VLBI properties of Mrk~421.
Other properties of this source, such as the low VLBI core brightness temperature,
also point to a low Lorentz factor on parsec scales (see Paper I), and these low Lorentz factors on
parsec scales seem to be a common feature of TeV blazars (Piner \& Edwards 2004), and possibly of HBLs 
in general (Giroletti et al. 2004b).
The VLBA observations of these sources then provide a valuable outer boundary condition
for models of high-energy emission through
shocks in the inner jet;
namely, that $\Gamma\sim2$ at projected distances of one parsec from the central black hole
($\approx$50 parsecs de-projected distance for an angle to the line-of-sight of 1$\arcdeg$).
After this paper was submitted, Gopal-Krishna, Dhurde, \& Wiita (2004) suggested
that an analysis
that takes into account the finite opening angle of the jet can
reproduce the slow apparent speeds in a jet that still has a high
Lorentz factor.  We plan in a future paper to undertake a
comparison of the Gopal-Krishna et al. (2004) model with our VLBI data,
to see if this model can explain the observed VLBI properties of the TeV
sources.

These dual-circular polarization observations also allowed us to study the polarization properties of
the Mrk~421 jet over an 11 month period starting several months after a prolonged gamma-ray high state.
The VLBI core and inner jet (component C7) had fractional polarizations of $\sim5\%$, and 
electric vector position angles (EVPAs) aligned with the jet axis.
Component C5 (at 1.5 mas from the core) had a higher fractional polarization of $\sim15\%$, and
an EVPA nearly orthogonal to the jet axis.
Significant variability was detected in the EVPA of component C6, which at two of the four epochs showed an
EVPA aligned with the jet axis, possibly a sign of propagating disturbances that are more easily visible
on the polarization images.
This raises the question of whether the
propagation of shocks in the jets of these HBL sources might be easier to track on VLBI polarization
images than on total intensity images alone.  

\acknowledgments
Part of the work described in this paper has been carried out at the Jet
Propulsion Laboratory, California Institute of Technology, under
contract with the National Aeronautics and Space Administration.
The National Radio Astronomy Observatory is a facility of the National Science Foundation operated
under cooperative agreement by Associated Universities, Inc.
This research has made use of
the NASA/IPAC Extragalactic Database (NED) which is operated by the Jet Propulsion Laboratory, California Institute of
Technology, under
contract with the National Aeronautics and Space Administration. 
This research has made use of the United States Naval Observatory (USNO) Radio Reference Frame Image Database (RRFID).
We thank Ken Kellermann for providing model component data from the 2~cm survey. 
This work was supported by the
National Science Foundation under Grant No. 0305475,
and by a Cottrell College Science Award from Research Corporation.


\begin{references}

Aller, H.~D. 1970, ApJ, 161, 19

Bennett, C.~L., et al. 2003, ApJS, 148, 1

Edwards, P.~G., Piner, B.~G., Unwin, S.~C., Wehrle, A.~E., Murphy, D.~W.,
Hirabayashi, H., \& Fujisawa, K. 2000,
in Astrophysical Phenomena Revealed by Space VLBI, ed. H. Hirabayashi, P.~G. Edwards, \&
D.~W. Murphy (Sagamihara:ISAS), 235

Edwards, P.~G., \& Piner, B.~G. 2002, ApJ, 579, L67

Gaidos, J.~A., et al. 1996, Nature, 383, 319

Georganopoulos, M., \& Kazanas, D. 2003, ApJ, 594, L27

Ghisellini, G., Tavecchio, F., \& Chiaberge, M. 2004, preprint (astro-ph/0406093) 

Giroletti, M., et al. 2004a, ApJ, 600, 127

Giroletti, M., Giovannini, G., Taylor, G.~B., \& Falomo, R. 2004b, preprint (astro-ph/0408213)

Gopal-Krishna, Dhurde, S., \& Wiita, P.~J. 2004, ApJ, 615, L81

Guetta, D., Ghisellini, G., Lazzati, D., \& Celotti, A. 2004, A\&A, 421, 877

Hirabayashi, H., et al. 1998, Science, 281, 1825, and erratum 282, 1995 

Homan, D.~C., Ojha, R., Wardle, J.~F.~C., Roberts, D.~H., Aller, M.~F., Aller, H.~D., \&
Hughes, P.~A. 2002, ApJ, 568, 99

Horan, D., \& Weekes, T.~C. 2004, New Astronomy Reviews, 48, 527

Jorstad, S.~G., Marscher, A.~P., Mattox, J.~R., Wehrle, A.~E., Bloom, S.~D., \&
Yurchenko, A.~V. 2001a, ApJS, 134, 181

Jorstad, S.~G., Marscher, A.~P., Mattox, J.~R., Aller, M.~F., Aller, H.~D.,
Wehrle, A.~E., \& Bloom, S.~D. 2001b, ApJ, 556, 738

Katarzy\'{n}ski, K., Sol, H., \& Kus, A. 2003, A\&A, 410, 101

Kellermann, K.~I., et al. 2004, ApJ, 609, 539

Konopelko, A., Mastichiadis, A., Kirk, J., De Jager, O.~C., \& Stecker, F.~W.
2003, ApJ, 597, 851

Kovalev, Y.~Y., Nizhelsky, N.~A., Kovalev, Y.~A., Berlin, A.~B., Zhekanis, G.~V., Mingaliev, M.~G., Bogdantsov, A.~V.
1999, A\&AS, 139, 545

Krawczynski, H., et al. 2001, ApJ, 559, 187

Krennrich, F., et al. 1999, ApJ, 511, 149

Krennrich, F.,  et al. 2001, ApJ, 560, L45

Krennrich, F., et al. 2002, ApJ, 575, L9

Lister, M.~L. 2001, ApJ, 562, 208

Lyutikov, M., Pariev, V.~I., \& Gabuzda, D.~C. 2004, preprint (astro-ph/0406144)

Maraschi, L., et al. 1999, ApJ, 526, L81

Marscher, A.~P., Jorstad, S.~G., Mattox, J.~R., \& Wehrle, A.~E. 2002, ApJ, 577, 85

Piner, B.~G., Unwin, S.~C., Wehrle, A.~E., Edwards, P.~G., Fey, A.~L., \& Kingham, K.~A. 
1999, ApJ, 525, 176 (Paper~I)

Piner, B.~G., Fey, A.~L., Mahmud, M., \& Gospodinova, K. 2003, BAAS, 203, 92.07

Piner, B.~G., \& Edwards, P.~G. 2004, ApJ, 600, 115

Pollack, L.~K., Taylor, G.~B., \& Zavala, R.~T. 2003, ApJ, 589, 733

Punch, M., et al. 1992, Nature, 358, 477

Readhead, A.~C.~S. 1994, ApJ, 426, 51

Shepherd, M.~C., Pearson, T.~J., \& Taylor, G.~B. 1994, BAAS, 26, 987

Spada, M., Ghisellini, G., Lazzati, D., \& Celotti, A. 2001, MNRAS, 325, 1559

Tanihata, C., Takahashi, T., Kataoka, J., \& Madejski, G.~M. 2003, ApJ, 584, 153

Tavecchio, F. 2004, preprint (astro-ph/0401590)

Wehrle, A.~E., Piner, B.~G., Unwin, S.~C., Zook, A.~C., Xu, W., Marscher, A.~P., Ter\"{a}sranta, H.,
\& Valtaoja, E. 2001, ApJS, 133, 297
                                                                                                                       
Zavala, R.~T., \& Taylor, G.~B. 2003, ApJ, 589, 126

Zavala, R.~T., \& Taylor, G.~B. 2004, ApJ, 612, 749

Zhang, F.J., \& B\aa\aa th, L.B. 1990, A\&A, 236, 47 (ZB90)

\end{references}
\end{document}